\documentclass[12pt]{iopart}

\usepackage[FIGTOPCAP,small]{subfigure}
\usepackage{graphicx}
\usepackage{dcolumn}
\usepackage{bm}

\usepackage{xcolor}

\newcommand{\bra}[1]{\left< #1 \right\vert}
\newcommand{\ket}[1]{\left\vert #1 \right>}
\newcommand{\pare}[1]{\left( #1 \right)}
\newcommand{\abs}[1]{\left\vert #1 \right\vert}
\newcommand{\cor}[1]{\left[ #1 \right]}

\begin{document}

\title[]{Generation of squeezed Schr\"{o}dinger cats in a tunable cavity filled with a Kerr medium }
\author{R de J Le\'on-Montiel$^{1}$ and H M Moya-Cessa$^{1}$}
\address{$^1$Instituto Nacional de Astrof\'{i}sica, \'{O}ptica y Electr\'{o}nica, Calle Luis Enrique Erro 1, Santa Mar\'{i}a Tonantzintla, Puebla CP 72840, Mexico}

\ead{robertoleonm@gmail.com}

\begin{abstract}
In this work we analyze the temporal dynamics of a system comprising an optical cavity filled with a nonlinear Kerr medium, whose frequency is allowed to change during time evolution. By exactly solving the corresponding time-dependent anharmonic-oscillator Hamiltonian, we demonstrate that squeezed coherent-state superpositions can be generated within the optical cavity. Moreover, we show that the squeezing degree of the produced states may be tuned by properly controlling the frequency shift of the cavity, a feature that might lead to interesting studies in the field of quantum state engineering.
\end{abstract}
\submitto{\JO}

\section{Introduction}
Generation of nonclassical states of light has been a subject of interest for many years \cite{vogel1993,parkins1993,parkins1995,brune1996}. In particular, the preparation of coherent-state superpositions \cite{yurke1986,schleich1991} have attracted special attention \cite{ourjoumtsev2006,neergaard-nielsen2006,wakui2007} due to their important role in many quantum information applications, such as quantum information processing \cite{braunstein_book}, quantum computation \cite{ralph2003}, and quantum-based precision measurements \cite{munro2002}. Interestingly, the special case of squeezed coherent-state superpositions has shown to be useful for many quantum information protocols, including quantum teleportation and single-qubit gates \cite{marek2008}, mainly because of its robustness against dissipative effects \cite{serafini2004}, and the large amplitudes that can be experimentally achieved \cite{ourjoumtsev2007}.

Different schemes for the generation of coherent-state superpositions have been proposed and realized over the years. Among these, we can mention schemes based on photon-number state and homodyne detection \cite{ourjoumtsev2007}, cavity QED \cite{raimond2001,biao2010,hermann2015}, two-photon subtraction or addition \cite{takahashi2008,marek2008_2,gerrits2010,wang2013}, and Kerr nonlinearities \cite{gerry1999,paris1999,jeong2005,he2009,song2013}. Among these proposals, interaction schemes based on the Kerr effect have been extensively criticized, mainly because Kerr nonlinearities are generally weak, thus requiring a long interaction time between the field and the nonlinear medium. Because in experiments we are dealing with open systems, during a long interaction time dissipation effects (such as decoherence) may destroy coherent quantum superpositions, making Kerr-based schemes seem unrealistic. One way to overcome this problem is by implementing schemes where strong Kerr nonlinearities are used \cite{schmidt1996,imamoglu1997}. Indeed, in the strong nonlinearity regime, the environment will affect the optical system during a \emph{shorter} interaction time, thus making the nonlinear interaction effect more dominant than decoherence. Another alternative is to use weak nonlinearities together with small initial amplitudes of the field, as it has been shown that in this regime decoherence effects are negligible \cite{paris1999}.
 
In this work we analyze a system, based on a tunable optical cavity filled with a nonlinear Kerr medium, that may be used for the preparation of squeezed coherent-state superpositions. We obtain a closed expression for an arbitrary initial state of the system and demonstrate that, under certain conditions, squeezed-state superpositions can be generated within the cavity. Furthermore, we show that the squeezing degree of the produced states can be easily tuned, a feature that may lead to interesting studies in the field of quantum state engineering.

This paper is organized as follows. In Section \ref{sect:model}, we present the theoretical model of the analyzed system. In Section \ref{sect:squeezed}, we describe a particular, exactly solvable, example in which squeezed-state superpositions, with tunable degree of squeezing, can be produced inside the cavity. In Section \ref{sect:wigner}, we analyze the nonclassical features of the generated states by evaluating their Wigner phase-space distribution. Finally, in Section \ref{sect:conclusions}, a summary of the results is provided.

\section{The model}\label{sect:model}

Let us consider a system comprising an optical cavity filled with a nonlinear Kerr medium. We assume that the cavity contains just one mode of the electromagnetic field and that its length, and therefore its frequency, can change during time evolution. This system could be experimentally realized by making use of a high-finesse piezo-driven tunable optical cavity \cite{mohle2013} filled with a low density atomic medium working under EIT conditions \cite{imamoglu1997}.

The dynamics of the described system is governed by the time-dependent Schr\"{o}dinger equation, with $\hbar = 1$,
\begin{equation}\label{schrodinger}
i\frac{\partial \ket{\psi\pare{t}}}{\partial t} = \hat{H}\pare{t}\ket{\psi\pare{t}},
\end{equation}
where the time-dependent anharmonic-oscillator Hamiltonian is given by \cite{milburn1986,milburn1986_2,paris1999}
\begin{equation}\label{hamiltonian}
\hat{H}\pare{t} = \frac{1}{2}\cor{\hat{p}^{2} + \Omega^{2}\pare{t}\hat{q}^{2}} + \chi\pare{\hat{a}^{\dagger}\hat{a}}^{2},
\end{equation}
with $\Omega\pare{t}$ being the time-dependent frequency of the oscillator, and $\chi$ the third order nonlinear susceptibility of the medium.

In principle, the way in which the $\hat{a}$ operator is written as
the sum of two squares need not be unique. We could have for instance the annihilation operator
\begin{equation}
\hat{a}_1\pare{t} = \frac{1}{\sqrt{2\Omega(t)}}\hat{q} + i\sqrt{\frac{\Omega(t)}{2}}\hat{p} ,
\end{equation}
or the annihilation operator defined at $t=0$ \cite{Recamier}
\begin{equation}
\hat{a}_2\pare{0} = \frac{1}{\sqrt{2\Omega(0)}}\hat{q} + i\sqrt{\frac{\Omega(0)}{2}}\hat{p} .
\end{equation}
Here, we take the operators $\hat{a}$ and $\hat{a}^{\dagger}$ to be defined as \cite{lewis1967}
\begin{equation}\label{a}
\hat{a}\pare{t} = \frac{1}{\sqrt{2}}\pare{\frac{\hat{q}}{\rho} + i\pare{\rho\hat{p} - \dot{\rho}\hat{q}}},
\end{equation}
\begin{equation}\label{a_dagger}
\hat{a}^{\dagger}\pare{t} = \frac{1}{\sqrt{2}}\pare{\frac{\hat{q}}{\rho} - i\pare{\rho\hat{p} - \dot{\rho}\hat{q}}},
\end{equation}
where $\rho$ obeys the Ermakov equation
\begin{equation}\label{ermakov}
\ddot{\rho} + \Omega^{2}\pare{t}\rho = \rho^{-3}.
\end{equation}
It is important to remark that we use definitions given by Eqs. (\ref{a}) and (\ref{a_dagger}), because it has been shown that these operators are, indeed, the corresponding generalized annihilation and creation operators of the time-dependent harmonic oscillator \cite{hartley1982,guasti2003_1,castanos2013}. Moreover, they obey the equation of motion for the Lewis-Ermakov invariant \cite{guasti2003_1}, namely
\begin{equation}\label{ermakov}
\frac{d\hat{a}}{dt}=\frac{i}{\rho^2(t)}[\hat{I},\hat{a}],
\end{equation}
where
\begin{equation}\label{ermakov}
\hat{I}=\frac{1}{2}\left(\left(\frac{\hat{q}}{\rho}\right)^2+\left(\rho\hat{p}-\dot{\rho}\hat{q}\right)^2\right),
\end{equation}
is the so-called Lewis-Ermakow invariant \cite{Ray,Reid,Haas,Pedrosa}.

Now, following the method introduced in Ref. \cite{guasti2003_2}, we can easily find that Eq. (\ref{schrodinger}) has a solution of the form
\begin{eqnarray}\label{psi_1}
\ket{\psi\pare{t}} &=& \hat{T}^{\dagger}\pare{t}\exp\cor{-i\chi\pare{\hat{q}^{2} + \hat{p}^{2} -1}^{2}/4} \nonumber \\
&  & \times \exp\cor{-i\pare{\hat{q}^{2} + \hat{p}^{2}}\int_{0}^{t}\omega\pare{t'}dt'/2}
\hat{T}\pare{0}\ket{\psi\pare{0}},
\end{eqnarray}
where $\omega\pare{t}=1/\rho^{2}$, and
\begin{equation}
\hat{T}\pare{t} = \exp\cor{i\frac{\mbox{ln}\rho}{2}\pare{\hat{q}\hat{p} + \hat{p}\hat{q}}}\exp\pare{-i\frac{\dot{\rho}}{2\rho}\hat{q}^{2}}.
\end{equation}

Equation (\ref{psi_1}) can be further simplified by considering the initial state of the system to be a coherent state, that is,
\begin{equation}\label{initial}
\ket{\psi\pare{0}} = \hat{T}\pare{0}\ket{\psi\pare{0}} = \ket{\alpha}.
\end{equation}
Then, by substituting Eq. (\ref{initial}) into Eq. (\ref{psi_1}), we can write the time evolution of the system as
\begin{eqnarray}\label{psi_2}
\ket{\psi\pare{t}} &=& \exp\pare{-\frac{\abs{\alpha}^{2}}{2}}\exp\pare{i\frac{\dot{\rho}}{2\rho^{3}}\hat{q}^{2}}\hat{S}\pare{r} \nonumber \\
& & \times \sum_{n=0}^{\infty}\frac{1}{\sqrt{n!}}\cor{\alpha\exp\pare{-i\int_{0}^{t}\omega\pare{t'}dt'}}^{n}
\exp\pare{-i\chi n^{2}t}\ket{n},
\end{eqnarray}
where $\ket{n}$ represents a number state, and the squeezing operator $\hat{S}\pare{r}$ is described by
\begin{equation}
\hat{S}\pare{r} = \exp\cor{\frac{r}{2}\pare{\hat{a}\hat{a} - \hat{a}^{\dagger}\hat{a}^{\dagger}}},
\end{equation}
with $r=-\mbox{ln}\rho$. Here $\hat{a}$ and $\hat{a}^{\dagger}$ stand for the annihilation and creation operators of the harmonic oscillator with constant frequency, respectively.

Equation (\ref{psi_2}) constitutes the main result of the present work. The importance of this result resides in the fact that it describes, without any approximation, the state of an electromagnetic field contained in a tunable cavity filled with a nonlinear medium. Furthermore, as we will show next, this result can be used, under certain conditions, to produce superpositions of coherent squeezed states with tunable degree of squeezing.

\section{Generation of squeezed coherent-state superpositions}\label{sect:squeezed}

\begin{figure}[t!]
    \begin{center}
       \includegraphics[width=10.0cm]{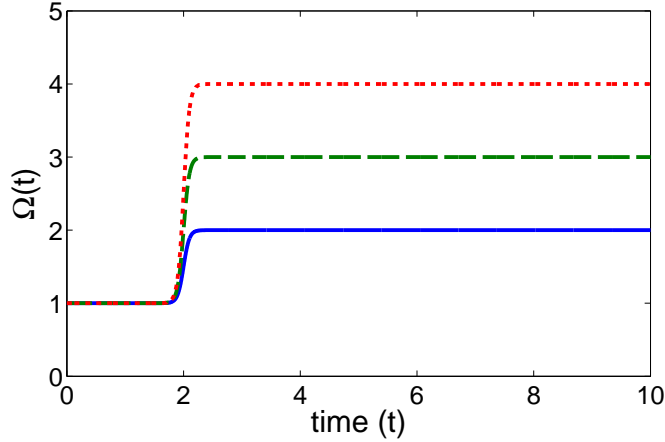}
    \end{center}
\caption{Frequency of the optical cavity, $\Omega\pare{t}$, as a function of time for $\omega_{i} = 1$, and different values of the final frequency: $\omega_{f} = 2$ (solid line), $\omega_{f} = 3$ (dashed line), and $\omega_{f} = 4$ (dotted line). We set $t_{s} = 2$ and $\epsilon = 10$.}
\label{fig:Omega}
\end{figure}

We now present a particular, exactly solvable, case for which squeezed-state superpositions can be obtained. To this end, we assume that during time evolution the length of the cavity can be suddenly modified. Indeed, any sudden change in the size of the cavity would result in a time-dependent frequency that may be modeled as a step-like function given by \cite{moya2003}
\begin{equation}\label{omega_step}
\Omega\pare{t} = \omega_{i}\cor{1 + \frac{\Delta}{2\omega_{i}}\pare{1 + \mbox{tanh}\cor{\epsilon\pare{t-t_{s}}}}},
\end{equation}
where $t_{s}$ is the time at which the frequency is changed, $\Delta = \omega_{f} - \omega_{i}$, with $\omega_{i}$ and $\omega_{f}$ being the initial and final frequencies of the cavity, and $\epsilon$ is a parameter that describes how fast the frequency is changed.

Figure \ref{fig:Omega} shows the frequency of the cavity as a function of time for $\omega_{i} = 1$, considering different values of the final frequency, $\omega_{f}$. Notice that the cavity's frequency changes to its final value in a rather short period of time. This frequency-shifting time can be controlled by modifying the parameter $\epsilon$, which in the limit $\epsilon \rightarrow \infty$ describes an ideal step function.

\begin{figure}[t!]
    \begin{center}
       \includegraphics[width=10.0cm]{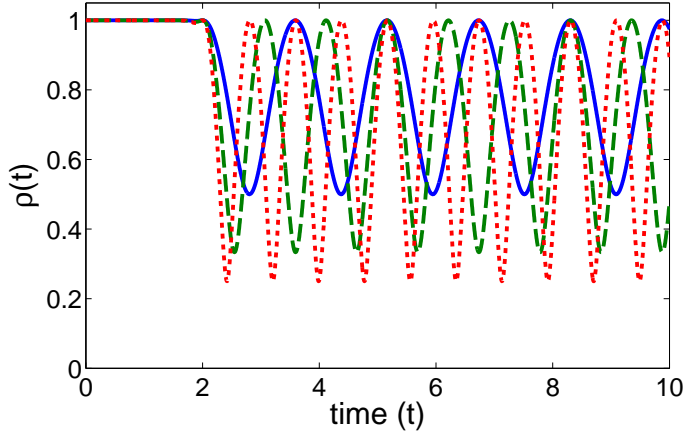}
    \end{center}
\caption{Solution to the Ermakov equation, $\rho\pare{t}$, as a function of time for $\omega_{i} = 1$, and different values of the final frequency: $\omega_{f} = 2$ (solid line), $\omega_{f} = 3$ (dashed line), and $\omega_{f} = 4$ (dotted line). We set $t_{s} = 2$ and $\epsilon = 10$.}
\label{fig:rho}
\end{figure}

Using Eq. (\ref{omega_step}), we can readily find that the Ermakov equation [Eq. (\ref{ermakov})] for this particular form of $\Omega\pare{t}$ has a solution given by \cite{moya2003}
\begin{equation}
\rho\pare{t} = \frac{1}{\sqrt{2}}\left\{1 + \frac{\omega_{i}^{2}}{\Omega^{2}\pare{t}} + \pare{1 - \frac{\omega_{i}^{2}}{\Omega^{2}\pare{t}}}\right.
\left.\cos\cor{2\int_{t_{s}}^{t}\Omega\pare{t'}dt'}\right\}^{1/2}.
\end{equation}
Figure \ref{fig:rho} shows $\rho\pare{t}$ for the same values given in Fig. \ref{fig:Omega}. We observe that $\rho\pare{t}$ exhibits an oscillatory behavior with local minima at specific values of time $t_{\mbox{\scriptsize{min}}}$. In the following, we will show that it is precisely at those moments in time when the electromagnetic field contained in the cavity behaves as a superposition of squeezed coherent states.

We start by noting that for local minima $\dot{\rho}\pare{t_{\mbox{\scriptsize{min}}}} = 0$, which means that at time $t_{\mbox{\scriptsize{min}}}$, Eq. (\ref{psi_2}) takes the form
\begin{eqnarray}\label{psi_3}
\ket{\psi\pare{t_{\mbox{\scriptsize{min}}}}} &=& \exp\pare{-\frac{\abs{\alpha}^{2}}{2}}\hat{S}\pare{r_{\mbox{\scriptsize{min}}}}\sum_{n=0}^{\infty}\frac{1}{\sqrt{n!}}\cor{\alpha\exp\pare{-i\int_{0}^{t_{\mbox{\scriptsize{min}}}}\omega\pare{t'}dt'}}^{n} \nonumber \\
& & \times \exp\pare{-i\chi n^{2}t_{\mbox{\scriptsize{min}}}}\ket{n},
\end{eqnarray}
with $r_{\mbox{\scriptsize{min}}} = -\mbox{ln}\cor{\rho\pare{t_{\mbox{\scriptsize{min}}}}}$.

Then, we proceed by assuming that the nonlinear coefficient $\chi$ satisfies
\begin{equation}\label{condition}
\chi t_{\mbox{\scriptsize{min}}} = \pi/2,
\end{equation}
which is a condition that can be reached by properly controlling the nonlinear susceptibility of the medium inside the cavity. Indeed, it has been shown that full control of the nonlinear susceptibility may be possible by making use of nonlinear effects, such as electromagnetically induced transparency \cite{schmidt1996,imamoglu1997}. Finally, by substituting Eq. (\ref{condition}) into Eq. (\ref{psi_3}), we can write the state of the system as
\begin{equation}\label{final_state}
\ket{\psi\pare{t_{\mbox{\scriptsize{min}}}}} = \frac{1-i}{2}\ket{\tilde{\alpha};r_{\mbox{\scriptsize{min}}}} + \frac{1+i}{2}\ket{- \tilde{\alpha};r_{\mbox{\scriptsize{min}}}},
\end{equation}
with $\tilde{\alpha} = \alpha\exp\cor{-i\int_{0}^{t_{\mbox{\scriptsize{min}}}}\omega\pare{t'}dt'}$, and
\begin{eqnarray}\label{squeezed state}
\ket{\tilde{\alpha};r_{\mbox{\scriptsize{min}}}} &=& \mu^{-1/2}\exp\cor{-\abs{\beta}^{2}/2 + \nu\beta^{2}/(2\mu)} \nonumber \\
& & \times \sum_{n=0}^{\infty}\frac{1}{\sqrt{n!}}\cor{\nu/(2\mu)}^{n/2}H_{n}\pare{\beta/\sqrt{2\mu\nu}}\ket{n},
\end{eqnarray}
where $\beta = \mu\tilde{\alpha} + \nu\tilde{\alpha}^{*}$, $\mu = \mbox{cosh}\pare{r_{\mbox{\scriptsize{min}}}}$, $\nu = \mbox{sinh}\pare{r_{\mbox{\scriptsize{min}}}}$, and $H_{n}\pare{x}$ is the Hermite polynomial of order $n$. We can observe from Eq. (\ref{final_state}) that a superposition of squeezed states, with squeezing parameter $r_{\mbox{\scriptsize{min}}}$, can be generated in the cavity at periodic times $t_{\mbox{\scriptsize{min}}}$, which are defined by the points where the time derivative of $\rho\pare{t}$ vanishes.

To verify the nonclassical behavior of the produced states, in the next section, we analyze the properties of the quantum state given in Eq. (\ref{final_state}) by evaluating its corresponding Wigner phase-space distribution.


\section{Wigner phase-space distribution}\label{sect:wigner}

To compute the Wigner function of the quantum state described by Eq. (\ref{final_state}), we make use of the Wigner's series representation \cite{moya2006} to write
\begin{equation}\label{wigner}
W\pare{Q,P} = \frac{1}{\pi}\sum_{k=0}^{\infty}\pare{-1}^{k}\abs{\bra{k}\hat{D}^{\dagger}\pare{\gamma}\ket{\psi\pare{t_{\mbox{\scriptsize{min}}}}}}^{2},
\end{equation}
where $\gamma = \pare{Q+iP}/\sqrt{2}$, and $\hat{D}\pare{\gamma} = \exp\pare{\gamma\hat{a}^{\dagger} - \gamma^{*}\hat{a}}$ is the well-known displacement operator \cite{glauber1963}.

By substituting Eq. (\ref{final_state}) into Eq. (\ref{wigner}), we find that the Wigner function of the squeezed-state superposition [Eq. (\ref{final_state})] is given by
\begin{eqnarray}\label{wigner_fin}
W\pare{Q,P} &=& N\mu^{-1} \left[f f^{*}\mathcal{H}\pare{\beta_{+},\beta_{+}} + f g^{*}\mathcal{H}\pare{\beta_{+},\beta_{-}} + g f^{*}\mathcal{H}\pare{\beta_{-},\beta_{+}}\right. \nonumber \\
& & \left. +\; g g^{*}\mathcal{H}\pare{\beta_{-},\beta_{-}} \right],
\end{eqnarray}
with
\begin{equation}
f = (1-i)\exp\cor{-\abs{\beta_{+}}^{2}/2 + \nu\beta_{+}^{2}/(2\mu) },
\end{equation}
and
\begin{equation}
g = \;(1+i)\exp\pare{\xi^{*}\tilde{\alpha} - \xi\tilde{\alpha}^{*}}
\exp\cor{-\abs{\beta_{-}}^{2}/2 + \nu\beta_{-}^{2}/(2\mu) },
\end{equation}
where the coefficients $N$, $\beta_{\pm}$ and $\xi$ are defined as
\begin{equation}
N = \frac{1}{4\pi}\abs{\exp\pare{\frac{\xi\tilde{\alpha}^{*} - \xi^{*}\tilde{\alpha}}{2}}}^{2},
\end{equation}
\begin{equation}
\beta_{\pm} = \mu\pare{\pm\tilde{\alpha} + \xi} + \nu\pare{\pm\tilde{\alpha} + \xi}^{*},
\end{equation}
\begin{equation}
\xi = -\mu\gamma - \nu\gamma^{*},
\end{equation}
and the function $\mathcal{H}(x,y)$ has the form
\begin{eqnarray}
\mathcal{H}(x,y) &=& \frac{\mu}{\sqrt{\mu^{2} - \nu^{2}}}\exp\cor{-\frac{x}{\mu^{2}}\pare{y^{*} + \frac{\nu}{2\mu}x} } \nonumber \\
& & \times\exp\cor{-\frac{\nu}{2\mu\pare{\mu^{2}-\nu^{2}}}\pare{y^{*}+\frac{\nu}{\mu}x }^{2} }.
\end{eqnarray}

\begin{figure}[t!]
    \begin{center}
       \subfigure[]{\includegraphics[width=10.0cm]{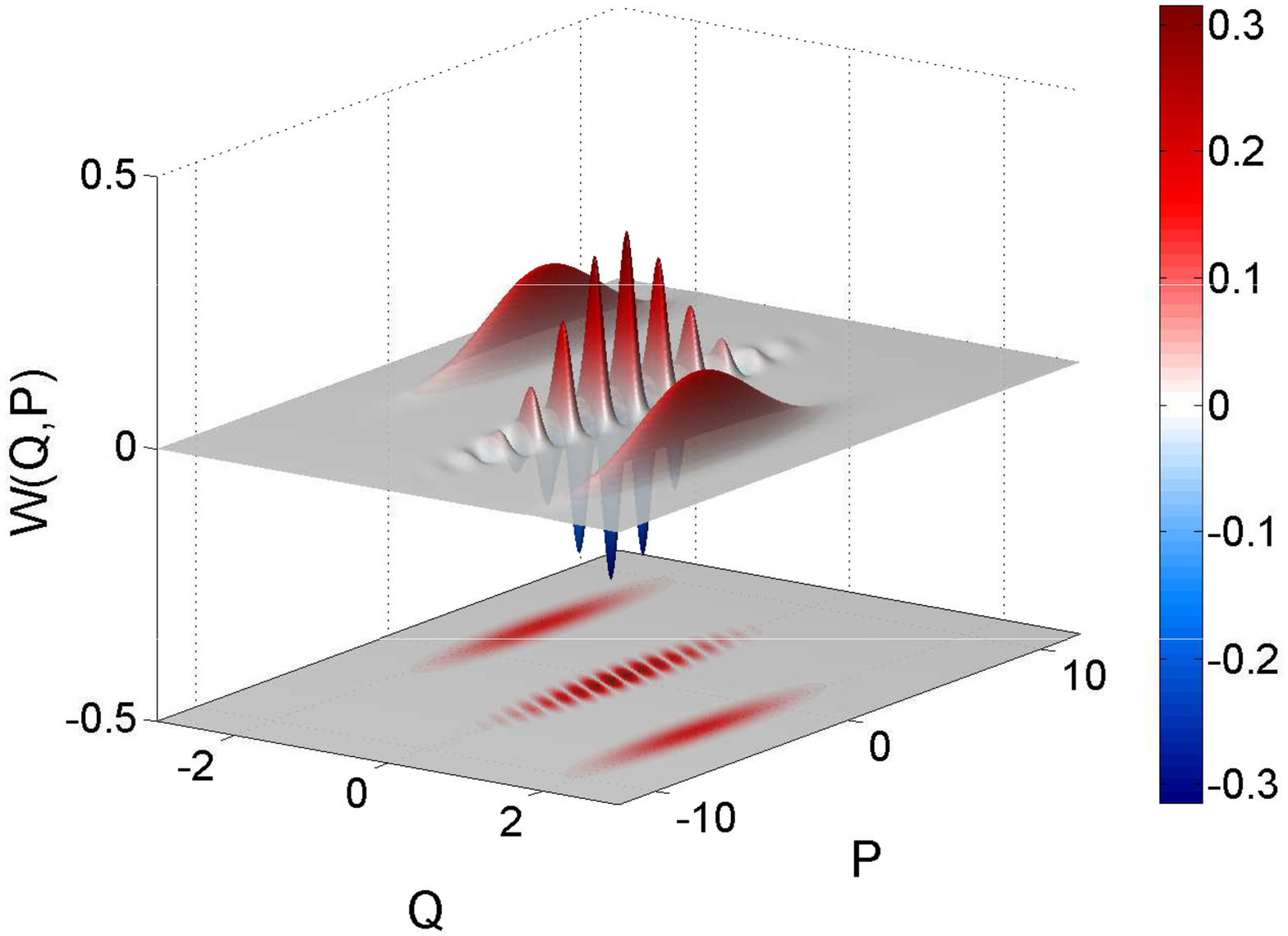}}
       \subfigure[]{\includegraphics[width=10.0cm]{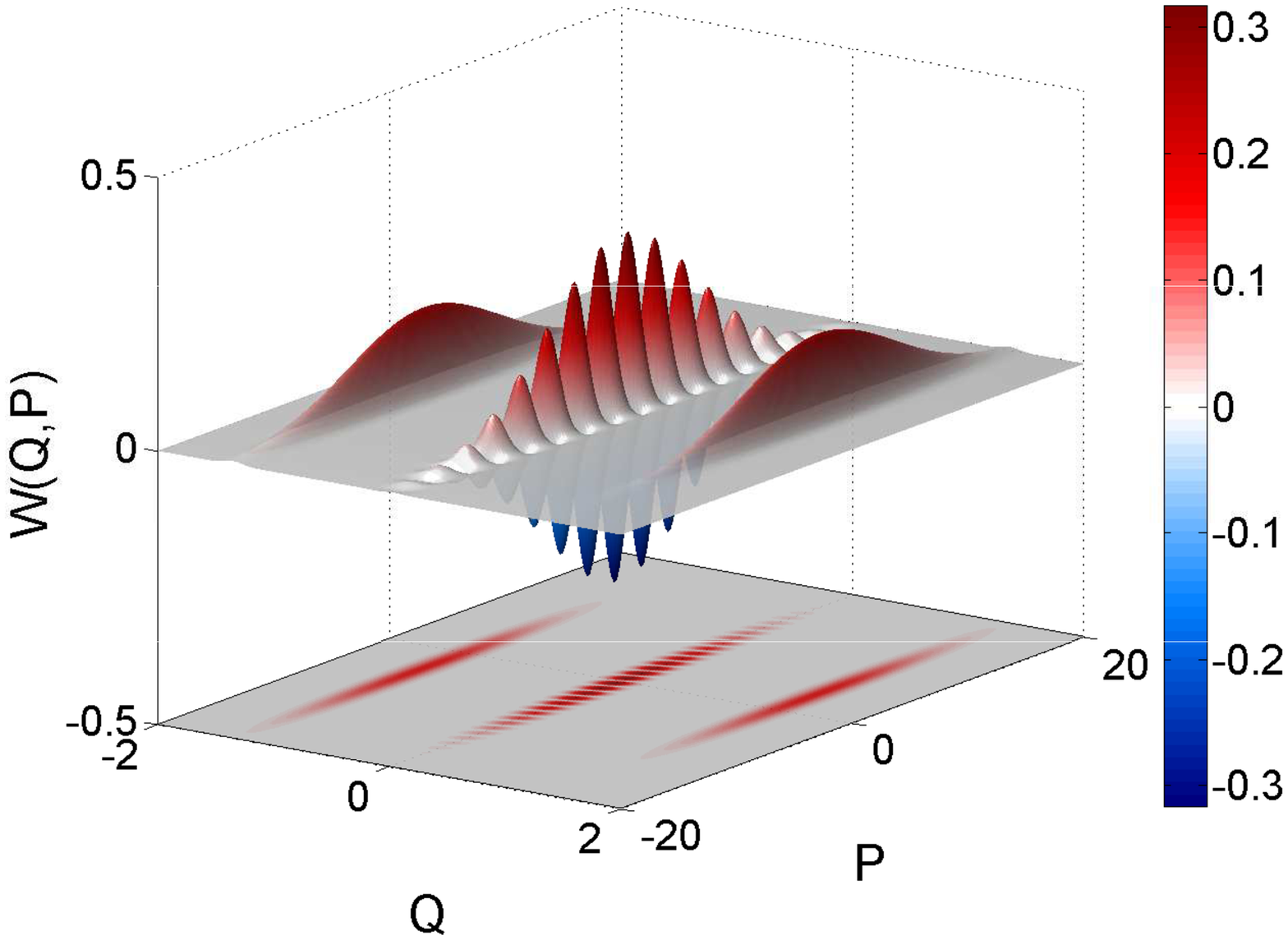}}
    \end{center}
\caption{Wigner function of the squeezed-state superposition given in Eq. (\ref{final_state}), with $\alpha = 3$, $\omega_{i} = 1$, and final frequency: (a) $\omega_{f} = 2$, and (b) $\omega_{f} = 4$. Parameters of the step-like function, $\Omega\pare{t}$, are set to $t_{s} = 2$ and $\epsilon = 10$.  }
\label{fig:wigner}
\end{figure}

Figure \ref{fig:wigner} shows the Wigner function of squeezed-state superpositions for $\omega_{i} = 1$ and final frequencies $\omega_{f} = 2$ [Fig. \ref{fig:wigner}(a)] and $\omega_{f}=4$ [Fig. \ref{fig:wigner}(b)]. Notice that, as predicted by Eq. (\ref{final_state}), an initially injected coherent state will be transformed by the action of the nonlinear material, and the tunable optical cavity, into a superposition of two squeezed coherent states, whose squeezing degree depends on the final frequency of the optical cavity; a larger frequency would produce a higher squeezing of the coherent states. This result implies that, indeed, the analyzed nonlinear optical system may be used to produce coherent superpositions of squeezed states with a tunable degree of squeezing.

\section{Conclusions}\label{sect:conclusions}

In this work, we have shown that a tunable optical cavity filled with a Kerr medium can be used for the generation of nonclassical states of light. We have provided an exact solution to the nonlinear time-dependent anharmonic-oscillator Hamiltonian and, using these results, we have found an optical system that may be used for direct generation of squeezed-state superpositions, whose degree of squeezing can be tuned by properly modifying the frequency shift of the optical cavity. Finally, we would like to remark that the analyzed system could be readily implemented by using presently available technology, namely a high-finesse piezo-driven tunable cavity filled with a low density atomic medium working under EIT conditions. This configuration may exhibit strong Kerr nonlinearities, which could help overcoming dissipation effects, such as decoherence, that are unavoidably present in realistic systems.

\ack

R.J.L.-M. acknowledges postdoctoral financial support from INAOE.

\end{document}